\documentclass[aps,prd,groupedaddress]{revtex4}
\usepackage{graphicx}
\usepackage{epsfig}
\usepackage{amsfonts}
\usepackage{amssymb}
\usepackage{epsf}
\usepackage{color}
\newcommand{\insertplot}[5]{\begin{figure}
 \hfill\hbox to 0.05in{\vbox to #5in{\vfill
 \inputplot{#1}{#4}{#5}}\hfill}
 \hfill\vspace{-.1in}
 \caption{#2}\label{#3}
 \end{figure}}
\newcommand{\inputplot}[3]{
 \special{ps: plotfile #1}
}

\newcounter{fig}

\newcommand{\ee}{\end{equation}}
\newcommand{\eea}{\end{eqnarray}}
\newcommand{\be}{\begin{equation}}
\newcommand{\bea}{\begin{eqnarray}}

\begin{document}

\title{(Un)balanced holographic superconductors with \\ electric and
spin motive force coupling}
\author{Nath\'alia P. Aprile}
\email[]{n.aprile@ifsc.usp.br}
\affiliation{Instituto de F\'isica de S\~ao Carlos, Universidade de S\~ao Paulo, S\~ao Carlos, S\~ao Paulo 13560-970, Brazil}
\email[]{b.hartmann@ucl.ac.uk}
\author{Betti Hartmann}
\email[]{b.hartmann@ucl.ac.uk}
\affiliation{Department of Mathematics, University College London, Gower Street, London, WC1E 6BT, UK}
\author{Jutta Kunz}
\email[]{jutta.kunz@uni-oldenburg.de}
\affiliation{Institute of Physics, University of Oldenburg, D-26111 Oldenburg, Germany}

\date{\today}

\begin{abstract}
We study  holographic phase transitions in (2+1) dimensions that possess interacting phases which result from a direct coupling between the two U(1) gauge fields. This can be interpreted as a non-minimal interaction between the electric and spin motive forces of the dual model. We first present a new analytical solution of the Einstein-Maxwell equations that describes a black hole with charge non-equivalent to the sum of the asymptotic charges of the two U(1) gauge fields and briefly discuss formation of uncharged scalar hair on this solution. We then study the formation of charged scalar hair on an uncharged black hole background and discuss the dual description of balanced as well as unbalanced superconductors. 

\end{abstract}

\maketitle

\section{Introduction}

Phase transitions (PT) can be grouped into universality classes which implies that systems with different microscopic descriptions can reveal the same asymptotic behavior for a given set of physical quantities. 
In recent years \cite{horowitz2011introduction,hartnoll2008holographic,hartnoll2008building,cai2015introduction} the Anti-de Sitter/Conformal Field Theory (AdS/CFT) correspondence has been used extensively  to find a dual description of diverse phase transitions, 
e.g. conductor/superconductor and insulator/superconductor phase transitions, using gravity theories in AdS space-time. 
When the PT is driven by thermal fluctuations, a natural tool on the gravity side are black holes which possess (in the classical description) the analogue of a temperature. 
Moreover, thermodynamical properties of black holes can be determined,
and in asymptotically AdS (aAdS) their specific heat is typically positive and signals thermodynamical stability. 
Note that this is very different in asymptotically flat space-time,
where it is well known that black holes have normally negative specific heat and would hence evaporate. 
The order parameter (or condensate) forming at the PT is then described by the formation of ``hair'' on the black hole \cite{gubser2008breaking}.

It has to be emphasized here that the main motivation for the construction of holographic superconductors has been high temperature superconductivity, 
i.e. the description of superconductors whose critical temperature lies well above the usual values for so-called conventional superconductors. 
Likely, in the former case the pairing mechanism between the electrons which leads to superconductivity is the result of a strong correlation and is not  phonon-mediated \cite{sachdev2011quantum}.

The range of materials that currently exhibit superconductivity is quite large:  
in addition to the observation of conventional superconductors, superconductivity has been observed in ferromagnetic materials, in organic conductors, in heavy fermion systems and in strontium ruthenate systems. 
This provides strong arguments for the existence of more exotic types of superconductivity. 
Hence, new types of phase transitions have to be considered in its modeling, an example being the addition of the interaction of the spins of the electrons. 

There exist superconducting ferromagnetic materials that can become superconducting when a magnetic field is applied. 
This mechanism allows the two species involved in the pairing phenomena to have distinct Fermi surfaces, which means that the two fermionic species can contribute to unbalanced populations or unbalanced chemical potentials in the superconducting phase. 
This process is called {\it inhomogeneous superconductivity} and exists e.g. in superconducting materials when doped \cite{casalbuoni2004inhomogeneous} and even in Quantum Chromodynamics (QCD). 

When the coupling between the external magnetic field and the conducting electrons is sufficiently strong, the electron pairing is broken and the system undergoes a first order PT from the superconducting to the normal phase. 
The limit at which this PT appears is called  the Chandrasekhar-Clogston (CC) bound \cite{clogston1962upper}. 
However, inhomogeneous superconductivity can lead to the spontaneous breaking of symmetries in the ground state 
and appropriate values of the difference between the chemical potentials allow the emergence of an anisotropically modulated gap,
which can energetically favor the appearance of a new phase in relation to the superconducting state and the non-superconducting normal state, respectively, known as the LOFF phase \cite{larkin1965nonuniform,fulde1964superconductivity}.

In fact, the application of an external magnetic field to the superconducting material generates a coupling between the orbital angular momentum of the electrons and the field, often suppressing the Zeeman interaction also present, due to the interaction of the field with the electronic spin moment. 
If we want to look at inhomogeneous superconductivity we have to take into account the imbalance caused by the presence of the Zeeman effect on these electrons.
This can be done by observing the electronic population associated with the spins up and down, $(n_{\uparrow})$ and $( n_{\uparrow})$, respectively, or directly in the imbalance between their chemical potentials, with $(\mu_{\uparrow})$ for the spin up and $(\mu_{\downarrow})$ for the spin down.
Although conventional superconductors can present these imbalances, there are geometries in non-conventional superconducting materials, such as cuprates, for example, that can favor this configuration. 
In this case, precisely because superconductivity occurs in parallel planes when we apply an external field in the direction of these planes, a perpendicular current appears between the planes that inhibits angular coupling and allows the Zeeman effect to become dominant \cite{nataliapizani}. 

The LOFF phase can also appear in other physical systems such as ultracold atoms in the transition between the superconducting phase and the Bose-Einstein phase \cite{zwierlein2006fermionic} 
or in the color interaction in the formation of the quark-quark condensate at low temperatures and high densities in QCD. 
In this latter case, the LOFF phase can be induced by the difference of the chemical potentials of the quarks involved, but also through the difference between 
their masses \cite{casalbuoni2018lecture}. 

In this paper, we study a minimal model of a holographic conductor-superconductor PT which contains competing phases
and possesses an explicit, non-minimal interaction between two U(1) gauge symmetries in order to mimic a direct interaction between the electric and spin motive forces.
The interaction is motivated by models suggested to explain dark matter \cite{Arkani-Hamed:2008hhe}, but in contrast to the original case, we will have two spontaneously broken gauge symmetries. 
Since we are mainly interested in the dual description on the superconductor side, we believe that this is a valid approach. 
Moreover, as previously mentioned, high temperature superconductivity involves complex and strongly correlated electronic interaction mechanisms. 
In fact, it appears that in these superconducting systems there is the emergence of PTs associated with different order parameters, for example, we have the emergence of a magnetic order competing with a superconducting order in materials that undergo the ferromagnetic-superconducting transition. 
In this sense, we can see our model as a two-phase s-type superconducting material.
Allowing these two scalar fields, i.e. two order parameters associated with the dual operator on the boundary to interact makes our model a dual description of a  \textit{multi-band superconductor}. 
The non-homogeneity will result from introducing differences between the chemical potentials or the charge densities, respectively. 

In particular, we discuss the formation of scalar hair on fixed black hole space-time backgrounds, i.e. we work in the so-called {\it probe limit}. 
We discuss two distinct cases here, considering 1.~a charged black hole as the background geometry and 2.~an uncharged black hole as the background geometry. 
In the first case, the two U(1) gauge fields are completely fixed, while they are dynamical in the second case. 
Since the main point of this paper is to discuss the influence of the new interaction between the gauge fields, we will focus mainly on the second case of dynamical gauge fields.
But we will do a qualitative analysis of the charged black hole background case first and present a closed form solution.

\section{The setup}

In the following, we will study a simple field theoretical model that contains two complex valued scalar fields, both {\it a priori} gauged under a U(1) symmetry. 
Gravity is minimally coupled and we introduce a negative-valued cosmological constant. 
The action of this model hence reads~:
\begin{equation}
\label{action}
S=\frac{1}{16\pi G}\int {\rm d}^4 x \sqrt{-g} \left( R - 2 \Lambda + 16\pi G{\cal L}_{m} \right)  \ , 
\end{equation}
where $R$ is the Ricci scalar, $G$ denotes Newton's constant and $\Lambda=-3/\ell^2 < 0$ is the cosmological constant with $\ell$ the AdS radius.  
The matter Lagrangian ${\cal L}_{m}$ is given by~:
\begin{equation}
\label{eq:matter_lagrangian}
{\cal L}_{m}=-D_{\mu} \varphi (D^{\mu} \varphi)^*-\frac{1}{4} F_{\mu\nu} F^{\mu\nu}
-D_{\mu} \xi (D^{\mu} \xi)^*-\frac{1}{4} H_{\mu\nu} H^{\mu\nu}
-V(\varphi,\xi) + \frac{\epsilon}{2} F_{\mu\nu}H^{\mu\nu}
\end{equation} 
with the covariant derivatives $D_\mu\varphi=\partial_{\mu}\varphi-ie_1 A_{\mu}\varphi$,
$D_\mu\xi=\partial_{\mu}\xi-ie_2 a_{\mu}\xi$
and the field strength tensors $F_{\mu\nu}=\partial_\mu A_\nu-\partial_\nu A_\mu$, 
$H_{\mu\nu}=\partial_\mu a_\nu-\partial_\nu a_\mu$ of the two U(1) gauge potentials $A_{\mu}$, $a_{\mu}$ with coupling constants $e_1$ and $e_2$. 
The term proportional to $\epsilon$ is an interaction term discussed previously in the context of a possible dark matter sector \cite{Arkani-Hamed:2008hhe}.  
Here, $\varphi$ and $\xi$ are complex scalar fields  with potential
\begin{equation}
V(\varphi,\xi)=m_{\varphi}^2\varphi\varphi^* + m_{\xi}^2 \xi\xi^*  + \lambda \varphi\varphi^*  \xi\xi^*  \ .
\end{equation}
$m_{\varphi}$ and $m_{\xi}$ denote the masses of the $\varphi$ and $\xi$ fields, respectively, and $\lambda$ denotes
the coupling between these two fields. 

The equations resulting from the variation of the action with respect to the metric and matter fields then leads to a set of coupled non-linear differential equations which read
\begin{equation}
\label{eq:scalar}
\frac{1}{\sqrt{-g}} D_{\mu} \left(\sqrt{-g} D^{\mu} \varphi\right) = m_{\varphi}^2\varphi+ \lambda \varphi  \xi\xi^*  \ \ \ , \ \ \ 
\frac{1}{\sqrt{-g}} D_{\mu} \left(\sqrt{-g} D^{\mu} \xi\right) = m_{\xi}^2\xi+ \lambda \varphi \varphi^* \xi  \ \ , 
\end{equation}
\begin{equation}
\label{eq:gauge}
\frac{1}{\sqrt{-g}} \partial_{\mu} \left[\sqrt{-g} (F^{\mu\nu}  -\epsilon H^{\mu\nu} )  \right]  = i e_1 \left[ \varphi^* D^{\nu} \varphi - \varphi (D^{\nu} \varphi)^*\right] \ \ , \
\end{equation}
\begin{equation}
\label{eq:gauge2}
\frac{1}{\sqrt{-g}} \partial_{\mu} \left[\sqrt{-g} (H^{\mu\nu}  -\epsilon F^{\mu\nu} )  \right]  = i e_2 \left[ \xi^* D^{\nu} \xi - \xi (D^{\nu} \xi)^*\right]  \ \ , 
\end{equation}
\begin{equation}
\label{eq:einstein}
G_{\mu\nu}-\frac{3}{\ell^2}= 8\pi G T_{\mu\nu} \ \ , \ \     T_{\mu\nu}=g_{\mu\nu}   {\cal L}_{m}  - 2 \frac{\partial {\cal L}_{m}}{\partial g^{\mu\nu}}
\end{equation}
where $T_{\mu\nu}$ is the energy-momentum tensor. \\

We assume staticity of the solutions and that all fields depend only on the coordinate $r$ which can be interpreted as an energy scale with $r\rightarrow \infty$, i.e. the AdS boundary corresponding to the UV regime of the dual quantum field theory. 
The Ansatz for the fields in planar coordinates $t,r,x,y$ reads~:
\begin{equation}
A_{\mu} dx^{\mu} = A_{t}(r) dt \ \ , \ \    a_{\mu} dx^{\mu} = a_{t}(r)dt \ \ , \ \  \varphi=\varphi(r) \ \ , \ \ 
\xi=\xi(r) 
\end{equation}
and the metric is
\begin{equation}
\label{eq:metric}
ds^2 = -N \sigma^2 {\rm d} t^{2} + \frac{1}{N} {\rm d}r^2 + r^2 \left({\rm d}x^2 + {\rm d}y^2\right).
\end{equation}
with the two metric function $N(r)$ and $\sigma(r)$. 
Inserting these Ans\"atze into the equations of motion (\ref{eq:scalar}),  (\ref{eq:gauge}), (\ref{eq:gauge2}), (\ref{eq:einstein}), we obtain
\begin{equation}
\label{eq:eq1}
\varphi''=-\left(\frac{2}{r}+\frac{N'}{N}+\frac{\sigma'}{\sigma}  \right)\varphi'-\frac{e_1^2 A_t^2 \varphi }{N^2\sigma^2} + m_{\varphi}^2 \frac{\varphi}{N}
+ \frac{\lambda \varphi \xi^2}{N}  \ , 
\end{equation}
\begin{equation}
\label{eq:eq2}
\xi''= -\left(\frac{2}{r}+\frac{N'}{N} +\frac{\sigma'}{\sigma} \right)\xi'-\frac{e_2^2 a_t^2 \xi}{N^2\sigma^2} + m_{\xi}^2 \frac{\xi}{N}
+ \frac{\lambda \xi \varphi^2}{N}   \ , 
\end{equation}
\begin{equation}
\label{eq:eq3}
 A_{t}''=-\left(\frac{2}{r}  -\frac{\sigma'}{\sigma} \right) A_t' + \frac{2}{1-\epsilon^2}\left(\frac{e_1^2 A_t \varphi^2}{N} + \epsilon\frac{e_2^2 a_t \xi^2}{N}\right)      \ , 
\end{equation}
\begin{equation}
\label{eq:eq4}
 a_{t}''=-\left(\frac{2}{r} -\frac{\sigma'}{\sigma} \right)  a_t' + \frac{2}{1-\epsilon^2}\left(\frac{e_2^2 a_t \xi^2}{N} + \epsilon\frac{e_1^2 A_t \varphi^2}{N}\right)   \ ,
\end{equation}
\begin{equation}
\label{eq:eqE}
\frac{N'}{r} + \frac{N}{r^2} - \frac{3}{\ell^2}= 8\pi G T_t^t  \ \ \ , \ \ \  \frac{2\sigma' N}{r \sigma} = 8\pi G \left( T_r^r - T_t^t    \right)   \ .
\end{equation}
The form of these equations makes clear that the case $\epsilon^2 \geq  1$ would lead to a very different theory than the original model for $\epsilon=0$. 
We hence restrict ourselves to values of $\epsilon \in (-1:1)$ in this paper.

The equations of motion (\ref{eq:eq1}) - (\ref{eq:eq4}) have to be solved numerically.
In order to reduce the number of couplings in the model, we can use the invariance of the model under given rescalings. 
First of all, there exists an invariance under the following rescaling~:
\begin{equation}
\label{eq:scaling1}
r\rightarrow \chi r \ \ , \ \ m_{\varphi}\rightarrow \frac{m_{\varphi}}{\chi}  \ \ , \ \ 
 m_{\xi}\rightarrow \frac{m_{\xi}}{\chi}  \ \ , \ \   e_1 \rightarrow \frac{e_1}{\chi} \ \ , \ \ 
 e_2 \rightarrow \frac{e_2}{\chi}   \ \ , \ \   \lambda \rightarrow \frac{\lambda}{\sqrt{\chi}}  \ \ , \ \ \ell \rightarrow \chi \ell \ \ , 
\end{equation}
where $\chi$ is a real constant. We will use this in the following to set $\ell\equiv 1$. 
In addition there is an invariance under the rescaling of the scalar and gauge fields~:
\begin{equation}
\label{eq:scaling2}
\varphi \rightarrow \frac{\varphi}{\chi}  \ \ , \ \  \xi \rightarrow \frac{\xi}{\chi} \ \ , \ \ 
A_t \rightarrow \frac{A_t}{\chi}   \ \ , \ \  a_t \rightarrow \frac{a_t}{\chi} \ \ , \ \  \lambda\rightarrow \chi^2 \lambda \ \ , \ \  G\rightarrow  \chi^2 G \ \ , \ \ e_i \rightarrow \chi e_i \ , i=1,2 \ ,
\end{equation}
which we will use to fix the value of $e_1=1$.
Note that we will also use the abbreviation $\alpha^2 = 4\pi G$. 
We are thus left with six parameters: $\alpha$, $e_2$, $m_{\varphi}$, $m_{\xi}$, $\lambda$ and $\epsilon$.
\\

In the following we are interested
only in the probe limit, i.e. the limit of the formation of scalar hair on a (charged or uncharged) black hole background with fixed horizon at $r=r_h$ and without taking backreaction into account, i.e. setting $\alpha=0$.
In order to find explicit solutions to the equations of motion, we have to fix appropriate boundary conditions.
These boundary conditions depend on the two cases studied:
1. the background corresponds to a charged black hole or 2. the background corresponds to an uncharged black hole.
In the first case we will present a solution in closed form, whereas in the second case we will integrate the equations of motion numerically.

The second case contains the major set of results of this work. 
In this case, in order for the matter fields to be regular at the horizon $r_h$ we need to impose the boundary conditions:
\begin{equation}
A_t(r_h)=0 \ \  , \ \   \varphi'(r_h)=\left. 
\frac{m^{2}_{\varphi}\varphi+\lambda\varphi\xi^{2}}{N'}\right|_{r=r_h} 
\end{equation}
and
\begin{equation}
a_t(r_h)=0 \ \ , \ \  \xi'(r_h)= \left.\frac{m^{2}_{\xi}\xi+\lambda\xi\varphi^{2}}{N'}\right|_{r=r_h}  \ .
\end{equation}
On the conformal boundary, which is a 2-dimensional plane $\mathbb{R}^2$, the scalar fields should have the following behaviour~:
\begin{equation}
  \varphi(r\gg1)= \frac{\varphi_{-}}{r^{\lambda_{-}}}+\frac{\varphi_+}{r^{\lambda_{+}}}  \ \ , \ \
  \xi(r\gg1)= \frac{\xi_{-}}{r^{\tilde{\lambda}_{-}}}+\frac{\xi_+}{r^{\tilde{\lambda}_{+}}}  \ , 
\end{equation}
where
\begin{equation}
\lambda_{\pm}=\frac{3}{2}\pm \sqrt{\frac{9}{4}+ m_{\varphi}^2} \ \ , \ \  \tilde{\lambda}_{\pm}=
\frac{3}{2}\pm \sqrt{\frac{9}{4}+ m_{\xi}^2}  \ ,
\end{equation}
while the U(1) gauge fields fall-off as follows~:
\begin{equation}
 A_t(r\gg1)= \mu_1-\frac{\rho_1}{r} \ \ , \ \   a_t(r \gg1)= \mu_2-\frac{\rho_2}{r} \ \ .
 \end{equation} 
Here $\mu_1$ and  $\mu_2$ correspond to the chemical potentials with the difference between these potentials denoted 
by $\delta \mu := \mu_1 - \mu_2$, and  $\rho_1$ and $\rho_2$ correspond to the charge densities. 
For $\rho_1=\rho_2$, the superconductors are said to be {\it balanced}, while for $\rho_1 \neq \rho_2$, they are {\it unbalanced}.
The chemical potentials $\mu_1$ and $\mu_2$ associated to the two gauge fields $A_\mu$ and $a_\mu$, respectively, combine to effective chemical potentials of the two species present in the superconductor as follows  
\begin{equation}
    \mu_1= \frac{\mu_{\uparrow}+\mu_{\downarrow}}{2}  \ \ , \ \
 \mu_2=\frac{\mu_{\uparrow}+\mu_{\downarrow}}{2}  \ . 
 \end{equation}
In fact, one of the gauge fields represents the electrical part of the superconductivity,  while the other interacts with the spin linked to superconductivity through a Zeeman coupling.  
Note that in the literature often $\mu_2=\delta \mu$ corresponding to the chemical potential mismatch which is the contribution that at $T=0$ is  relevant for the formation of the LOFF phase.

The thermodynamical quantities of the black hole that we will use in the following are the temperature $T$ and the entropy ${\cal  S}$, which are given by
\begin{equation}
\label{eq:temperature_entropy}
T= \frac{\sigma(r_h) N'\vert_{r=r_h}  }{4\pi}    \ \ , \ \   {\cal  S}=\pi r_h^2   \ . 
\end{equation}
In our numerical construction, we have found it useful to fix the horizon $r_h$ and vary $\varphi_{+}$ and $\xi_{+}$, which also leads to
a variation of the charge densities. Often, however, it is useful to consider the solutions
for fixed charge density. Using the rescalings in the model, we can translate our numerical solutions into solutions that
have constant charge densities $\rho_{\varphi}^{*}=\rho_{\xi}^{*}=1$. The $*$ will denote the properties of the solution
for this choice in the following. We then have
\begin{equation}
T^{*}= \frac{T}{\sqrt{\rho_{\varphi}+\rho_{\xi}}}\ \ , \ \ \varphi^{*}= \frac{\varphi_{+}}{\sqrt{\rho_{\varphi}+\rho_{\xi}}} \ \ , \ \ \xi^{*}=\frac{\xi_{+}}{\sqrt{\rho_{\varphi}+\rho_{\xi}}} \ . 
\end{equation}

\section{Formation of scalar hair on charged black hole backgrounds}

We first consider charged black hole backgrounds. 
In fact, for $\varphi\equiv 0$, $\xi\equiv 0$, the Einstein-Maxwell equations (\ref{eq:gauge}), (\ref{eq:einstein}) have an explicit black hole solution that can be given in closed form. 
This reads
\begin{equation}
\label{eq:RN}
N=-\frac{M}{r}+\frac{\alpha^2}{r^2}\left(\rho_1^2+ \rho_2^2 - 2\epsilon \rho_1\rho_2\right) + r^2 \ \ , \ \    \sigma\equiv 1 \  \ , \  \  A_t(r)=\mu_1-\frac{\rho_1}{r} \ \ , \ \   a_t(r)= \mu_2-\frac{\rho_2}{r} \ \ ,
 \end{equation} 
where $\mu_1$ and $\mu_2$ correspond to the chemical potentials on the AdS boundary, while $\rho_1$ and $\rho_2$ denote the charge densities. 
$M$ is an integration constant that is related to the charge densities and the horizon radius $r_h$ of the black hole. 
The temperature of the black hole reads
\begin{equation}
T=\frac{1}{4\pi}\left[3 r_h - \frac{\alpha^2}{r_h^3}\left(\rho_1^2+ \rho_2^2 - 2\epsilon \rho_1\rho_2\right) \right] =
\frac{1}{4\pi}\left[3 r_h - \frac{\alpha^2}{r_h}\left(\mu_1^2+ \mu_2^2 - 2\epsilon \mu_1\mu_2\right) \right] 
\ .
\end{equation}
The temperature becomes zero for a black hole with horizon radius
\begin{equation}
\label{eq:rh_ex}
r_h=\left(\frac{\alpha^2}{3} \left(\rho_1^2+ \rho_2^2 - 2\epsilon \rho_1\rho_2\right) \right)^{1/4} =
\left(\frac{\alpha^2}{3} \left(\mu_1^2+ \mu_2^2 - 2\epsilon \mu_1\mu_2\right) \right)^{1/2} =
 \left(\frac{M}{4}\right)^{1/3} \ . 
\end{equation}
Thus, the new interaction will influence the radius of the black hole horizon and hence its entropy ${\cal S}$. 
At fixed gravitational back-reaction and charge densities (or chemical potentials, respectively), the entropy of the extremal black hole will decrease (increase) when increasing (decreasing) the coupling constant $\epsilon$.

In order to understand the scalar field condensation in the background of this modified RNAdS solution, we consider the space-time close to the horizon. 
Defining $R:=r-r_h$, the near-extremal, near-horizon geometry of the solution reads 
\begin{equation}
\label{eq:metricads2}
{\rm d}s^2 = - 6 R^2 {\rm d}t^2 + \frac{1}{6R^2} {\rm d}R^2 + r_h^2 \left({\rm d}x^2 + {\rm d}y^2\right)   \ \ , \ \ 
A_t(r)=\frac{\rho_1}{r_h^2}R  \ \ , \ \   a_t(r)=\frac{\rho_2}{r_h^2}R
\end{equation}
with $r_h$ given by (\ref{eq:rh_ex}). 
The geometry is, as expected,  $AdS_2 \times \mathbb{R}^2$. 
We can then insert this solution into the two scalar-field equations. 
Dropping the $\frac{2}{r}$ term in the pre-factor of $\varphi'$ and $\xi'$, respectively, as this can be neglected with respect to $N'/N$ here and denoting the derivative with respect to $R$ by a dot this gives~:
\begin{equation}
\ddot{\varphi} + \frac{2}{R} \dot{\varphi}  - \frac{m_{{\rm eff},\varphi}^2}{R^2} \varphi= 0   \ \ , \ \  \ddot{\xi} + \frac{2}{R} \dot{\xi}  - \frac{m_{{\rm eff},\xi}^2}{R^2} \varphi= 0
\end{equation}
with the effective masses 
\begin{equation}
\label{eq:masses}
m_{{\rm eff},\varphi}^2 = m_{\varphi}^2  + \frac{\lambda \xi_h^2}{6} - \frac{e_1^2 \rho_1^2}{12\alpha^2 (\rho_1^2 + \rho_2^2 - 2 \epsilon \rho_1 \rho_2)} \ \ , \ \   m_{{\rm eff},\xi}^2 = m_{\xi}^2  + \frac{\lambda \varphi_h^2}{6} - \frac{e_2^2 \rho_2^2}{12\alpha^2 (\rho_1^2 + \rho_2^2 - 2 \epsilon \rho_1 \rho_2)}   \ ,
\end{equation}
where the index $h$ denotes evaluation on the horizon. 
Now, in the $AdS_2$ near-horizon, near-extremal geometry, the effective masses should drop below the 2-dimensional Breitenlohner-Freedman (BF) bound \cite{Breitenlohner:1982bm,Breitenlohner:1982jf}, which is $m_{BF,2}^2=-1/4$. 
As (\ref{eq:masses}) demonstrates, positive values of the coupling constant $\lambda$, specifying the strength of the interaction of the scalar fields, then make condensation harder, while negative values make condensation easier. 
For the parameter $\epsilon$, which can have values only between $-1$ and $1$ and specifies the strength of the new interaction between the vector fields, a positive value makes condensation easier, while a negative value makes condensation harder.

\section{Formation of scalar hair on uncharged black hole backgrounds}

We now consider uncharged black hole backgrounds.
In this case, the black hole space-time is given by a solution to the vacuum Einstein equation including a cosmological constant.
It is thus the planar Schwarzschild-AdS (pSAdS) solution with metric functions
and Hawking temperature given by~:
\begin{equation}
\label{eq:metricSADS}
\sigma(r)\equiv 1   \ \ , \ \ N(r) = \frac{r^2}{\ell^2} - \frac{M}{r} =\frac{r^2}{\ell^2}\left(1-\frac{r_h^3}{r^3}\right) \ , \
T=\frac{3 r_h}{4\pi \ell^2}  \ .
\end{equation}
While the fixed space-time background is not influenced by the new interaction, the two dynamical gauge fields now interact directly and non-trivially with each other. 

We have solved the coupled gauge-scalar field equations in the background of the solution (\ref{eq:metricSADS}) using a collocation solver for boundary value problems \cite{Ascher}.
For our numerical construction, we have here reduced the large parameter space by choosing $m_{\varphi}^2=m_{\xi}^2=-2$. 
Both choices are well above the Breitenlohner-Freedman bound $m_{\rm BF}^2 = -9/4$ ensuring stability of the conformal boundary.  We have also
chosen $e_2=1$, leaving thus as free parameters $\alpha$, $\lambda$ and $\epsilon$. 
We will work in the probe limit in the following and hence set $\alpha=0$.
We have further set $\varphi_{-}=0$ and $\xi_{-}=0$ such that the respective fall-off of the scalar fields is 
\begin{equation}
\varphi(r\gg1)= \frac{\varphi_+}{r^2} \ \ , \ \  \xi(r\gg1)= \frac{\xi_+}{r^2}    \ . 
\end{equation}
Then $\varphi_+$ and  $\xi_+$ can be interpreted as the expectation values of operators of dimension two in the boundary quantum field theory. 
Concretely, we have computed the solutions by choosing values for $\varphi_+$ and $\xi_+$. 
This choice fixes the chemical potentials $\mu_1$ and $\mu_2$ as well as the charge densities $\rho_1$ and $\rho_2$.  
In order to study the conductor-superconductor phase transition at fixed chemical potentials we can use the scaling relations (\ref{eq:scaling2}). 

We would first like to address the competition between the phases due to the presence of the two order parameters associated with the system. 
Therefore we focus on making the system non-homogeneous by promoting the explicit difference between the charge densities, such that, for $\rho_1=\rho_2$, the superconductors are said to be \textit{balanced}, while for $\rho_1\neq \rho_2$, are said to be \textit{unbalanced}.

\begin{figure}[h!]
\centering
\vspace{-4cm}
\includegraphics[scale=0.50]{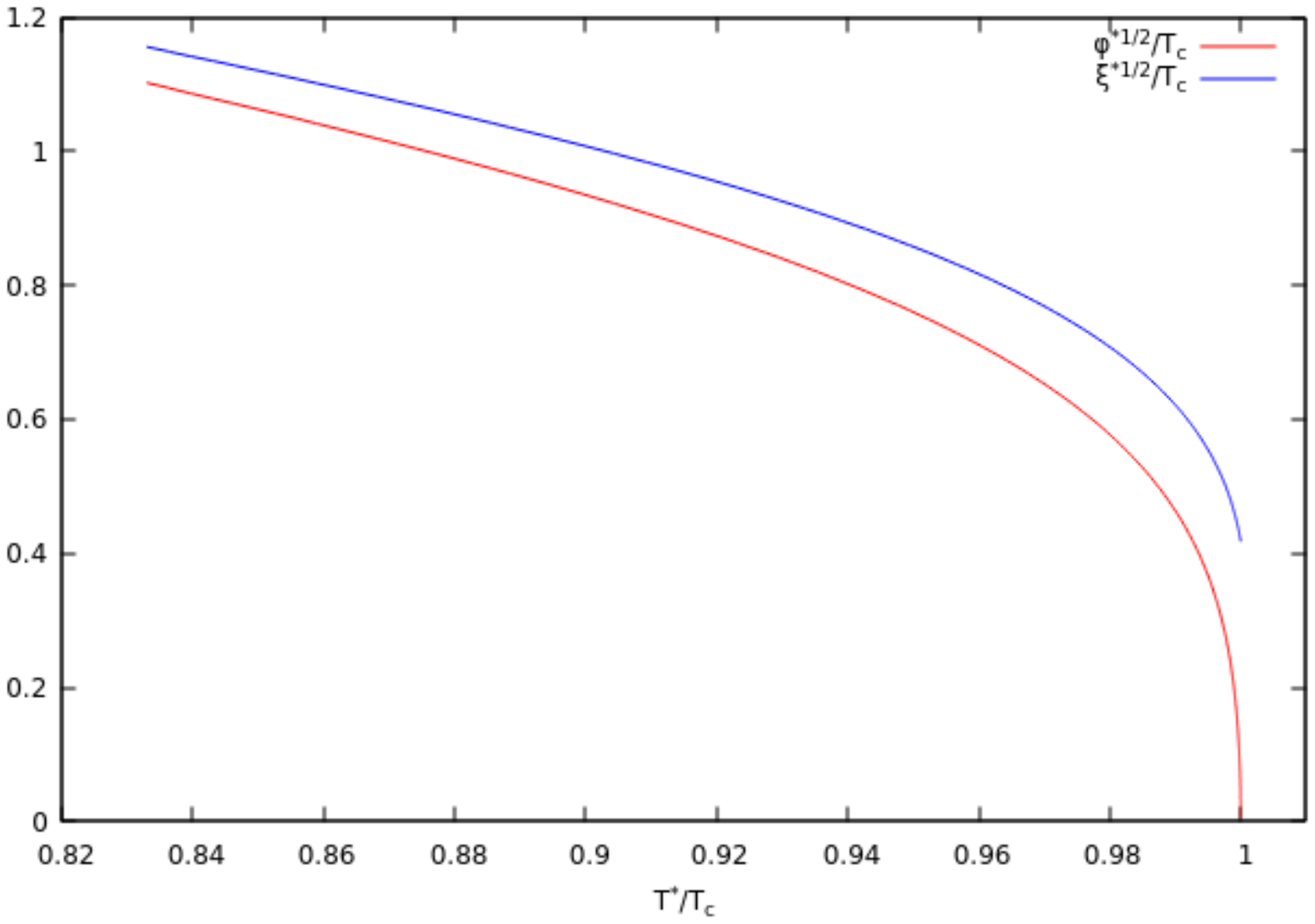}
\vspace{-5cm}
\caption{The dimensionless condensates $\varphi_+^{1/2}/T_c$ and $\xi_+^{1/2}/T_c$ associated with the scalar field $\varphi$ (red) and $\xi$ (blue), respectively, as functions of $T^*/T_c$, where $T_c$ is the temperature at which condensation sets in for $\varphi$. 
Here, $\epsilon=\lambda=0$, i.e. the two condensates do not interact. 
Note, that the curve in $\xi$ (blue) ends
at a temperature below its own condensation temperature, since 
the numerical parameter used to construct the curves is the value of the field $\varphi$.} 
\label{fig:condensade_lam0_eps_0}
\end{figure}

We have first - as a crosscheck of the validity of our numerical construction - reproduced known results \cite{Musso:2013ija, Musso:2013rva}.
In Fig.~\ref{fig:condensade_lam0_eps_0} we demonstrate that two distinct phases appear in the system due to the formation of scalar hair
on the pSAdS black hole associated with non-zero values for the scalar fields $\varphi$ (red) and $\xi$ (blue) near the horizon. 
The phases are non-interacting in this case as we have chosen  $\lambda=0$ and $\epsilon=0$ here. We have plotted the condensates versus 
$T^*/T_c$, where $T_c\approx 1.393$ is the critical temperature of $\varphi$. At this point, the value of $\xi_{+}$ was chosen to be $0.5$, i.e. we have modelled a situation in which the order parameter described by the $\xi$ field has already a considerable value when the condensation of the $\varphi$ field sets in.  
For $T > T_c$ it does not make sense to speak of a composite system as the system associated to $\varphi$ is above the condensation temperature. Hence the curve for $\xi$ stops here.

\subsection{Switching off the new interaction}

\begin{figure}[h]
\vspace{-3cm}
\includegraphics[scale=0.40]{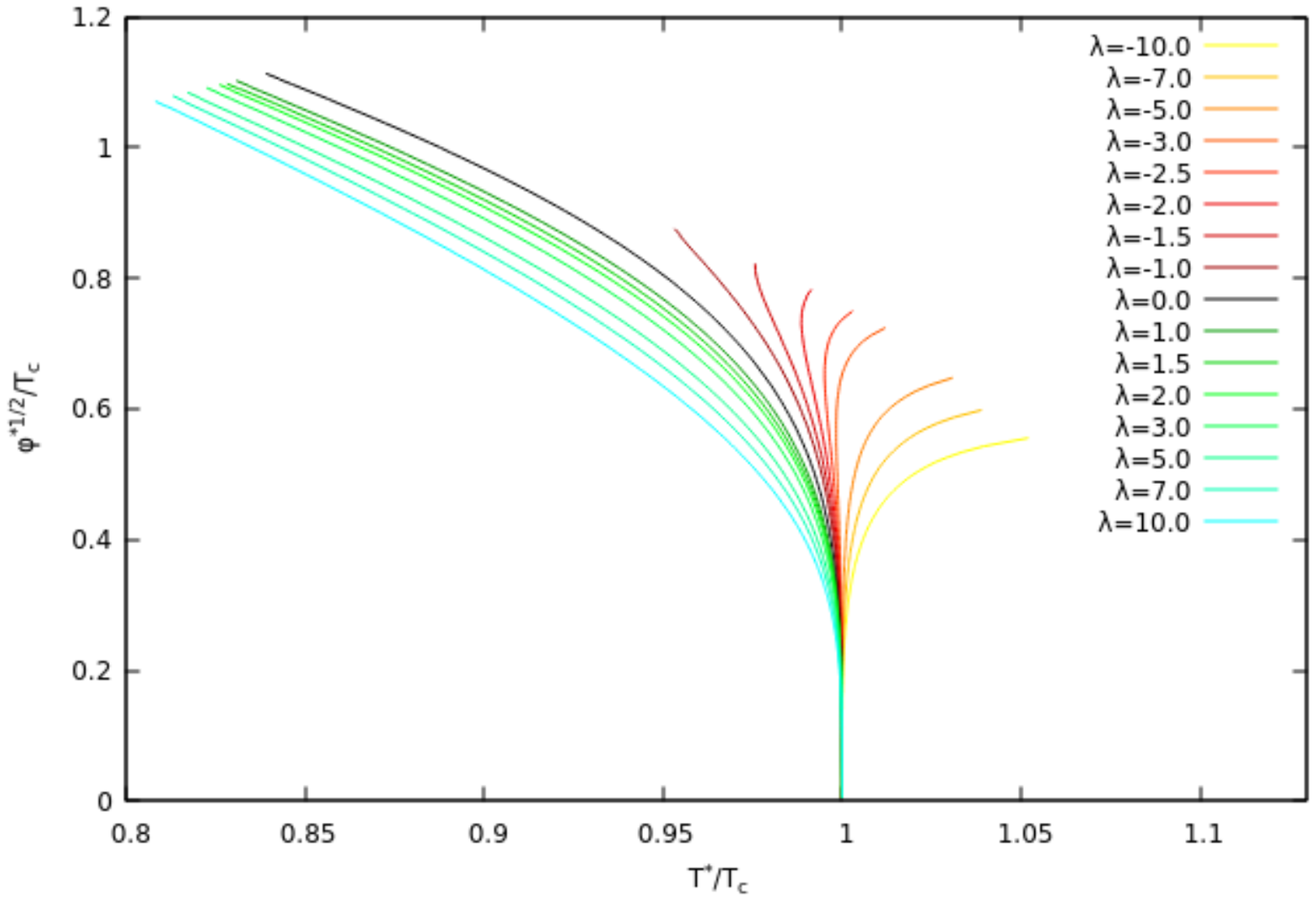}
\includegraphics[scale=0.40]{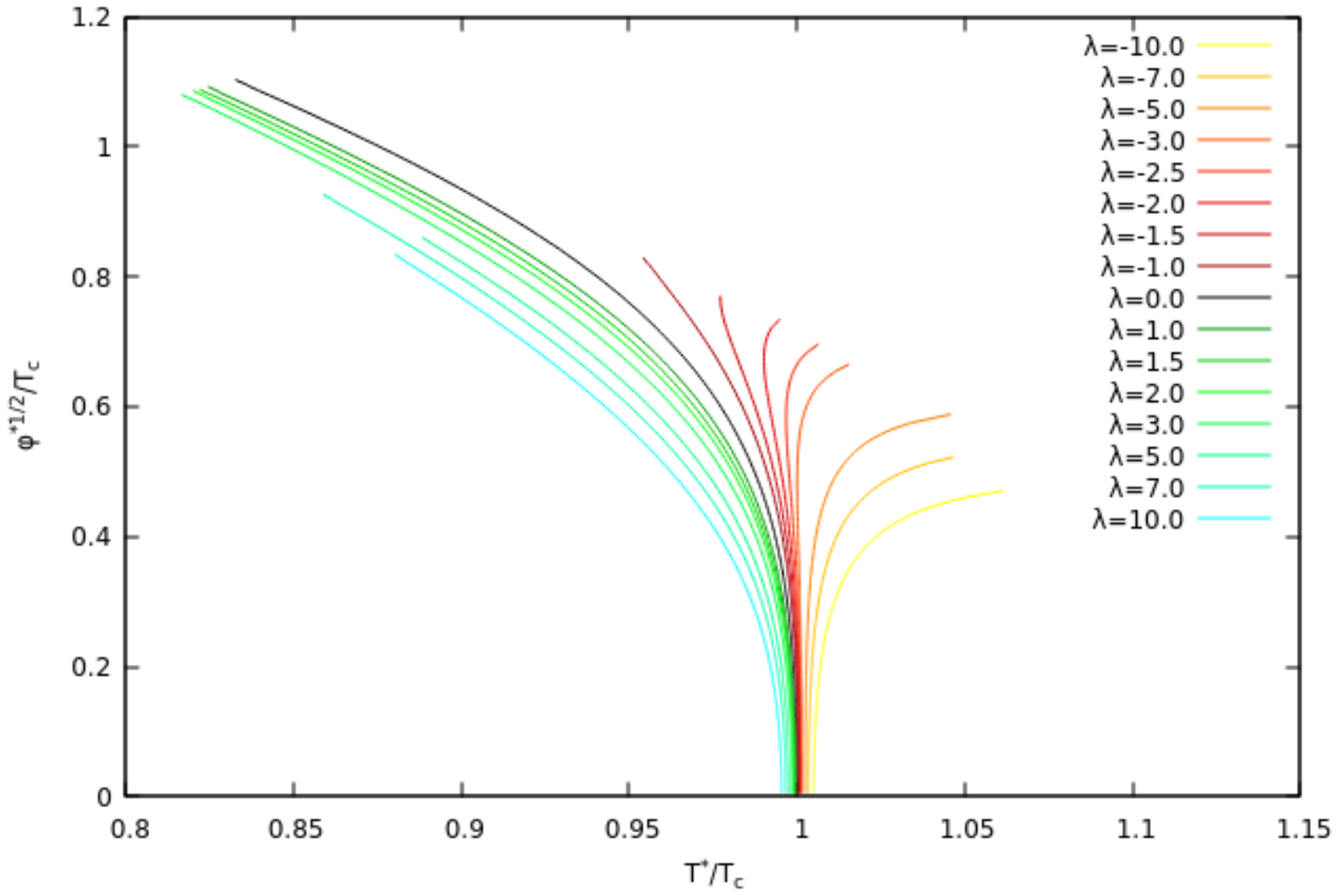}
 \vspace{-3cm}
 \caption{The condensate $\varphi_+^{1/2}/T_c$ as a function of $T^*/T_c$ for $\epsilon = 0$ and several values of $\lambda$. 
 We show the balanced case with $\rho_1=\rho_2$ (left) and the unbalanced case with $\rho_1\neq\rho_2$ (right).
 Note that the curves stop due to increased numerical difficulty to find a solution. \label{fig:condensade_balanced_unbalanced}}
 \end{figure}

In the following, we present the curves for $\varphi$ -- the curves for $\xi$ can be inferred from  Fig.~\ref{fig:condensade_lam0_eps_0}.
In order to introduce an interaction between the two phases, we have first followed \cite{Musso:2013ija, Musso:2013rva} and assumed an interaction of the scalar fields.
We will hence consider $\epsilon=0$ and study the qualitative change of the system when varying $\lambda$. 
This is shown in Fig.~\ref{fig:condensade_balanced_unbalanced} for the balanced (left) and the unbalanced (right) case, respectively.

As is obvious from this figure, the critical temperature $T_c$ of the balanced system remains unchanged when varying $\lambda$, while for the unbalanced case, the critical temperature changes its value as compared to the $\lambda=0$ case. 
Due to our choice of parameters, the balanced system is completely symmetric under the exchange of $\varphi$ and $\xi$ and hence the critical temperature should not change. 
The unbalanced case, on the other hand, shows that the condensation of the $\varphi$ field changes in the presence of the $\xi$ condensate.
In fact, increasing $\lambda$ from zero makes condensation harder - we have to decrease the temperature in order to find the onset of condensation, 
while choosing $\lambda < 0$ leads to an increase of the critical temperature, i.e. condensation becomes easier. 
Note that this is qualitatively similar to the results we have obtained in the case of a charged black hole background. 

Interestingly, we find that while for positive and vanishing value of $\lambda$,
the system shows a typical second order phase transition with condensate values increasing when lowering the temperature, this is different for negative $\lambda$. 
Here we find that the condensate increases for increasing temperature. This is, of course, not a typical conductor-superconductor phase transition,
but this type of behaviour appears in so-called Mott transitions in QCD (see e.g. \cite{MAO:Chiral_Mott_transition}). 

\begin{figure}[h]
\vspace{-3cm}
\includegraphics[scale=0.4]{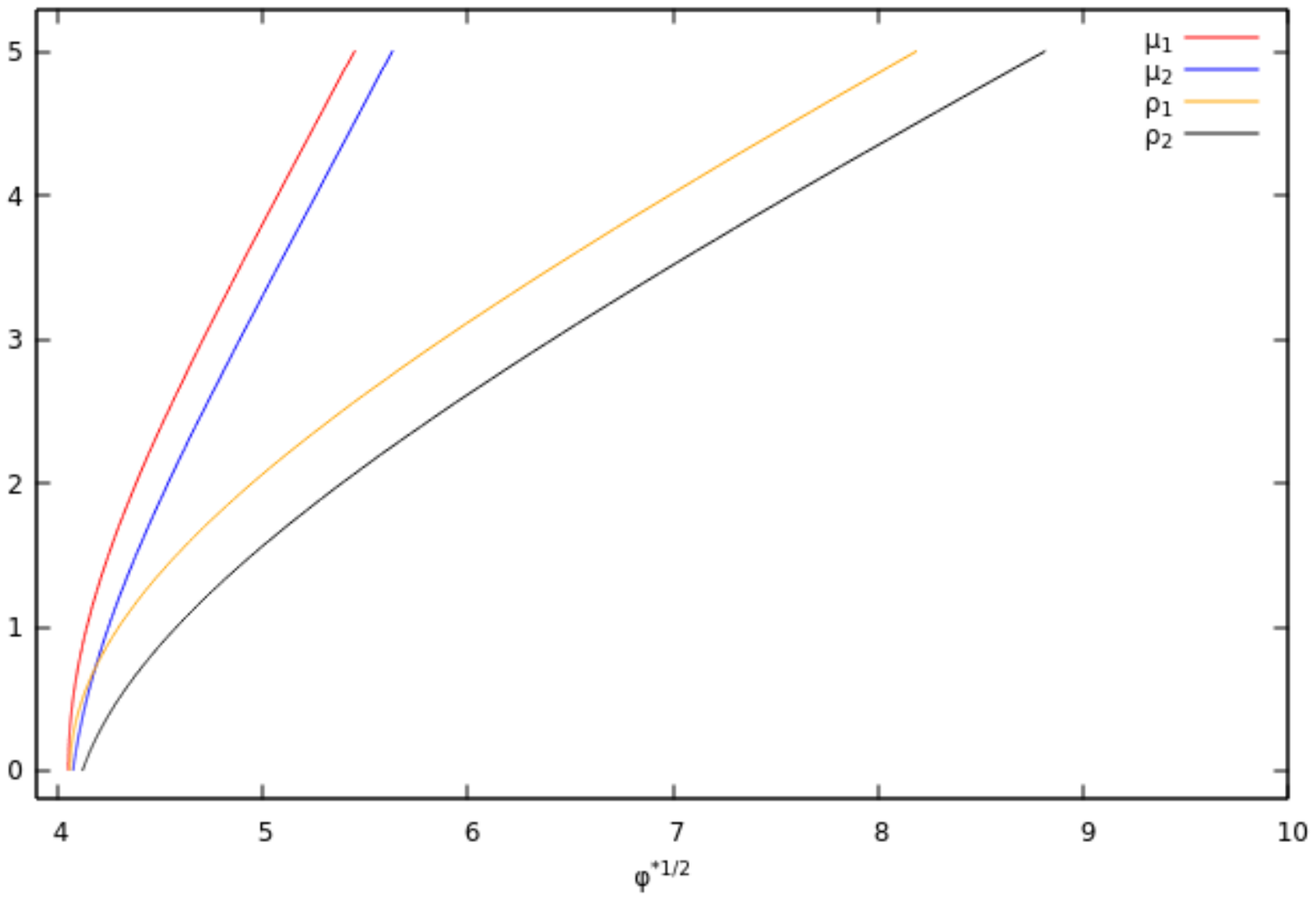}
\includegraphics[scale=0.4]{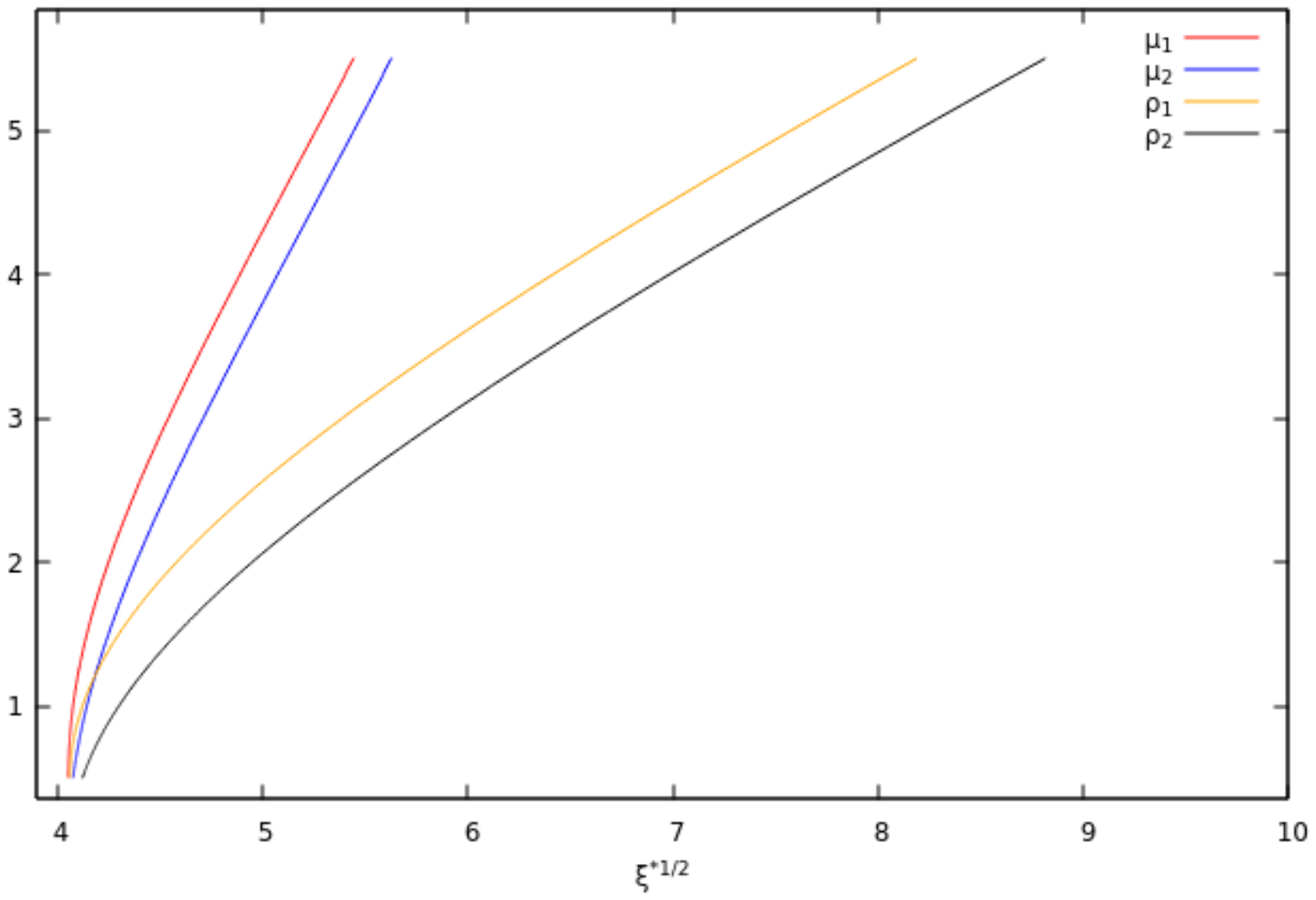}
\vspace{-3cm}
 \caption{We show the values of the chemical potentials $\mu_i$ and charge densities $\rho_i$, $i=1,2$, respectively, as functions of the condensate value $\varphi_+$ (left) and $\xi_+$ (right) for the unbalanced, non-interacting case.}
\label{fig:charge_chemical_non_interacting_unbalanced}
\end{figure}

We have also studied the charge densities and chemical potentials in the non-interacting, unbalanced case. 
They are shown in Fig.~\ref{fig:charge_chemical_non_interacting_unbalanced}.

All our results are in perfect agreement with those given in \cite{Musso:2013ija, Musso:2013rva}. 
We will now discuss the new case where the two phases interact via the U(1) gauge fields.

\subsection{Influence of the new interaction}

We now study the case $\epsilon\neq 0$. 
Here we first discuss the balanced case and then the unbalanced case. 
Finally, we comment on the possibility of the appearance of a LOFF phase in our model. 

\subsubsection{Balanced superconductors}

\begin{figure}[h]
\vspace{-3cm}
\includegraphics[scale=0.40]{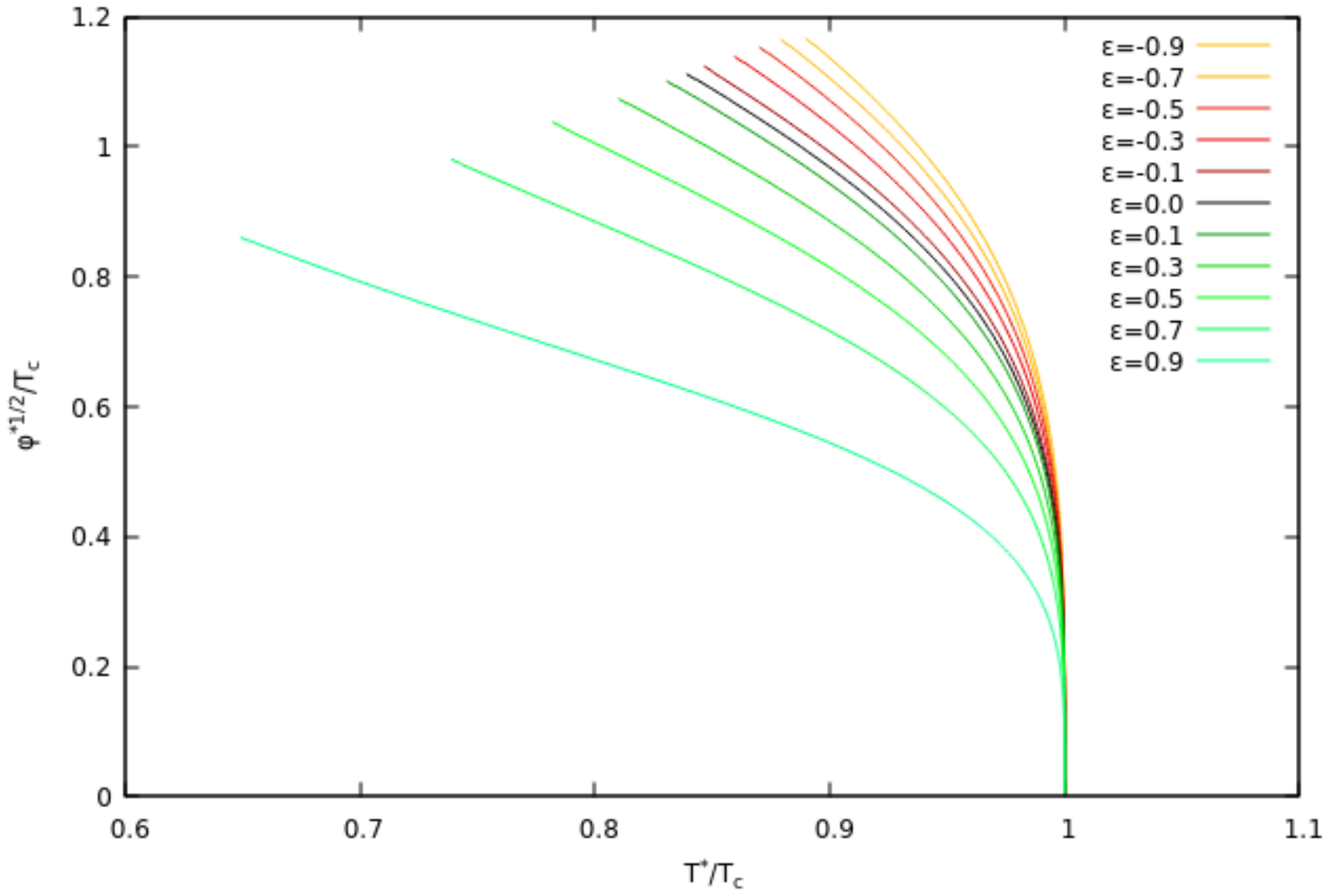}
\vspace{-3cm}
 \caption{The condensate $\varphi_+^{1/2}/T_c$ for a balanced superconductor as a function of $T/T_c$ for $\lambda = 0$ and several values of $\epsilon$. } 
 \label{fig:lab0epsilonbalanced}
 \end{figure}

We have first studied the case, where the interaction between the two phases
is solely mediated via the U(1) gauge fields. Our results for different
values of $\epsilon$ and $\lambda=0$ are shown in Fig.~\ref{fig:lab0epsilonbalanced}.

The black curve represents the non-interacting case, $\epsilon=0$. 
We find that when $\epsilon>0$ the value of the condensate decreases, while for $\epsilon<0$ it increases. 
Again, very similar and in accordance with our analysis in the case of a charged black hole background, positive (negative) $\epsilon$ makes condensation harder (easier). 

\begin{figure}[h]
\vspace{-3cm}
\includegraphics[scale=0.4]{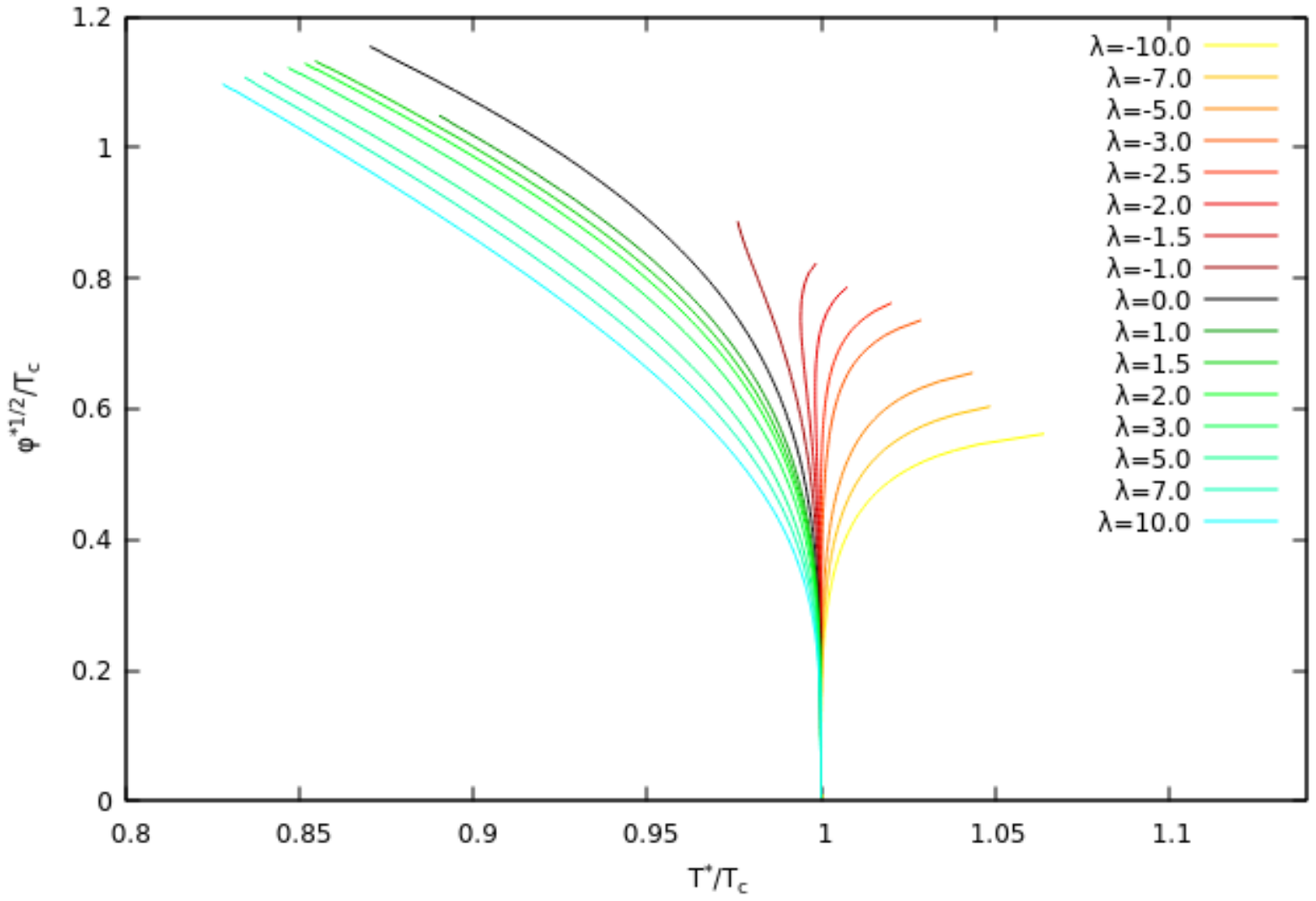}
\includegraphics[scale=0.4]{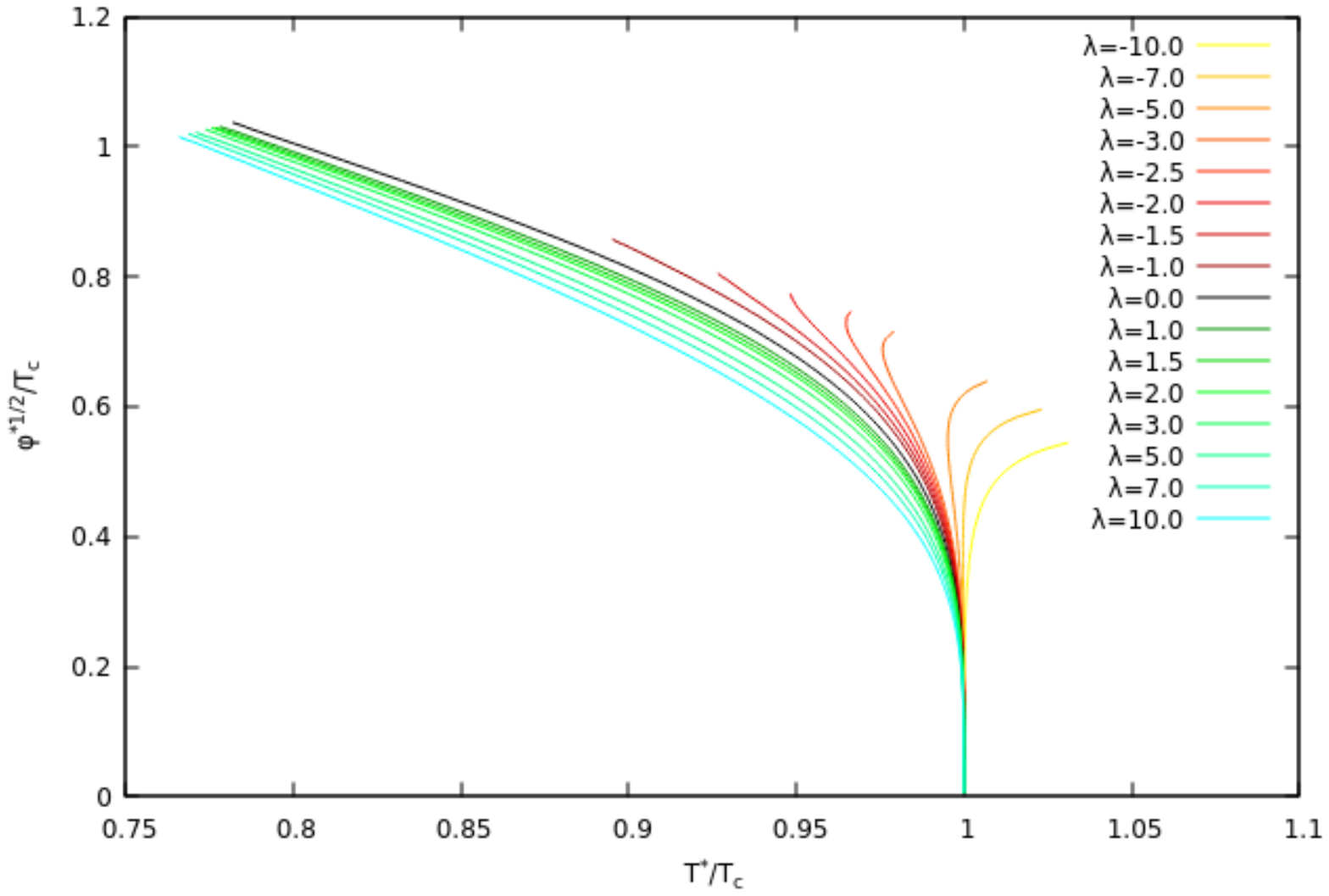}
\vspace{-3cm}
\caption{The condensate $\varphi_+^{1/2}/T_c$ for a balanced superconductor as a function of $T/T_c$  for several values of $\lambda$ and $\epsilon = -0.5$ (left) and $\epsilon = 0.5$ (right).}
\label{fig:balancedepsilon05}
 \end{figure}

\begin{figure}[h]
\vspace{-3cm}
\includegraphics[scale=0.4]{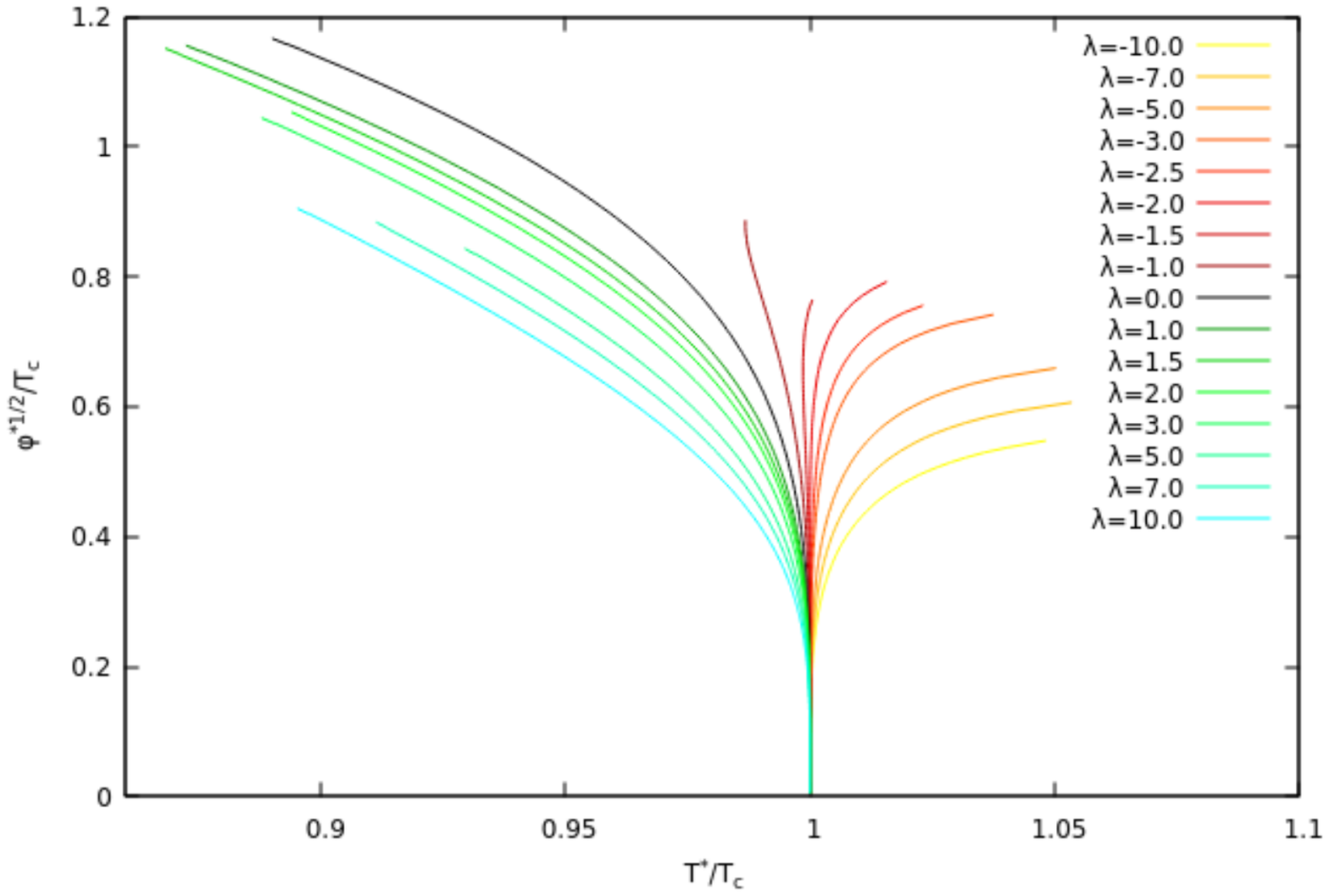}
\includegraphics[scale=0.4]{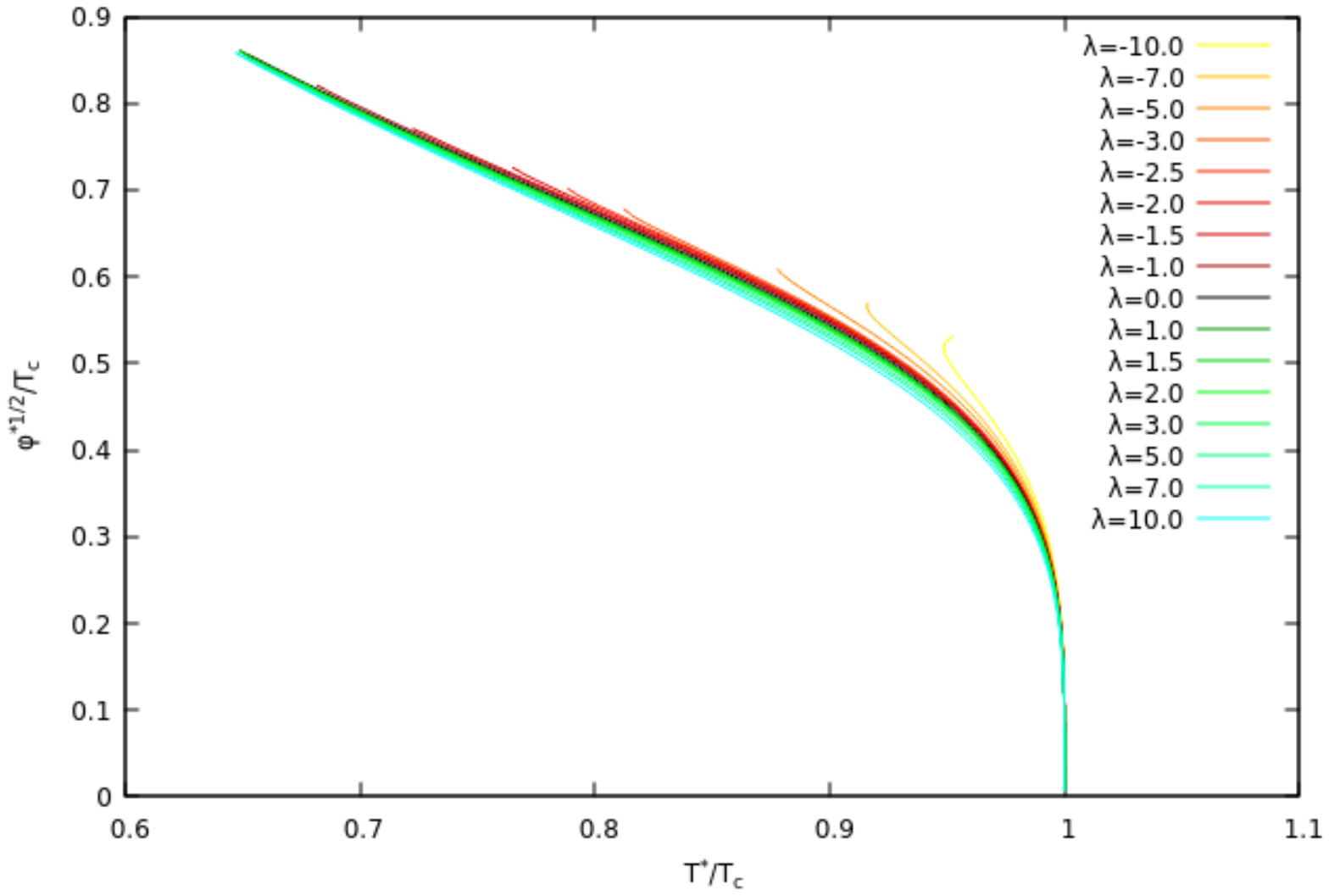}
\vspace{-3cm}
\caption{The condensate $\varphi_+^{1/2}/T_c$ for a balanced superconductor as a function of $T/T_c$  for several values of $\lambda$ and $\epsilon = -0.9$ (left) and $\epsilon = 0.9$ (right).}
\label{fig:balancedepsilon09}
 \end{figure}

Allowing next for both interactions in the system, we find that a competition can arise between them. 
Negative values of $\epsilon$ with $\lambda\neq 0$ do not cause much difference -
it seems that the scalar field coupling dominates the system. 
However, when $\epsilon$ is positive it starts to dominate the system, destroying the difference between the phases arising from the presence of $\lambda$. 
This is shown in 
Fig.~\ref{fig:balancedepsilon05} and 
Fig.~\ref{fig:balancedepsilon09} for $\epsilon=\pm 0.5$ and $\epsilon=\pm 0.9$, respectively. 

When varying $\epsilon$ we also find that the qualitative dependence of the charge density and the chemical potential vary little as compared to the $\epsilon =0$ case. 
Both are increasing functions of the condensation value of $\varphi_+$.

\subsubsection{Unbalanced superconductors}

\begin{figure}[h]
\vspace{-3cm}
\includegraphics[scale=0.4]{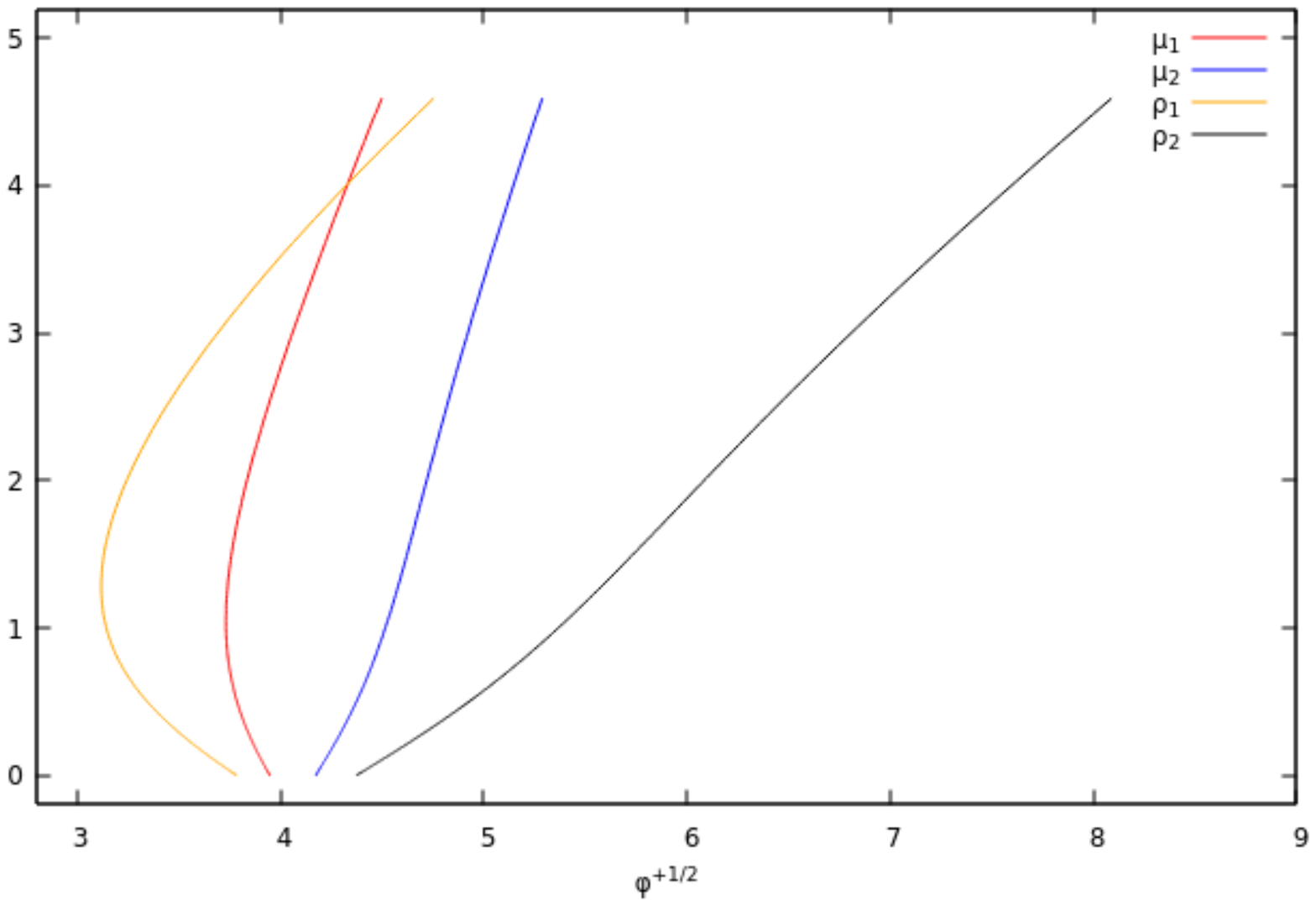}
\includegraphics[scale=0.4]{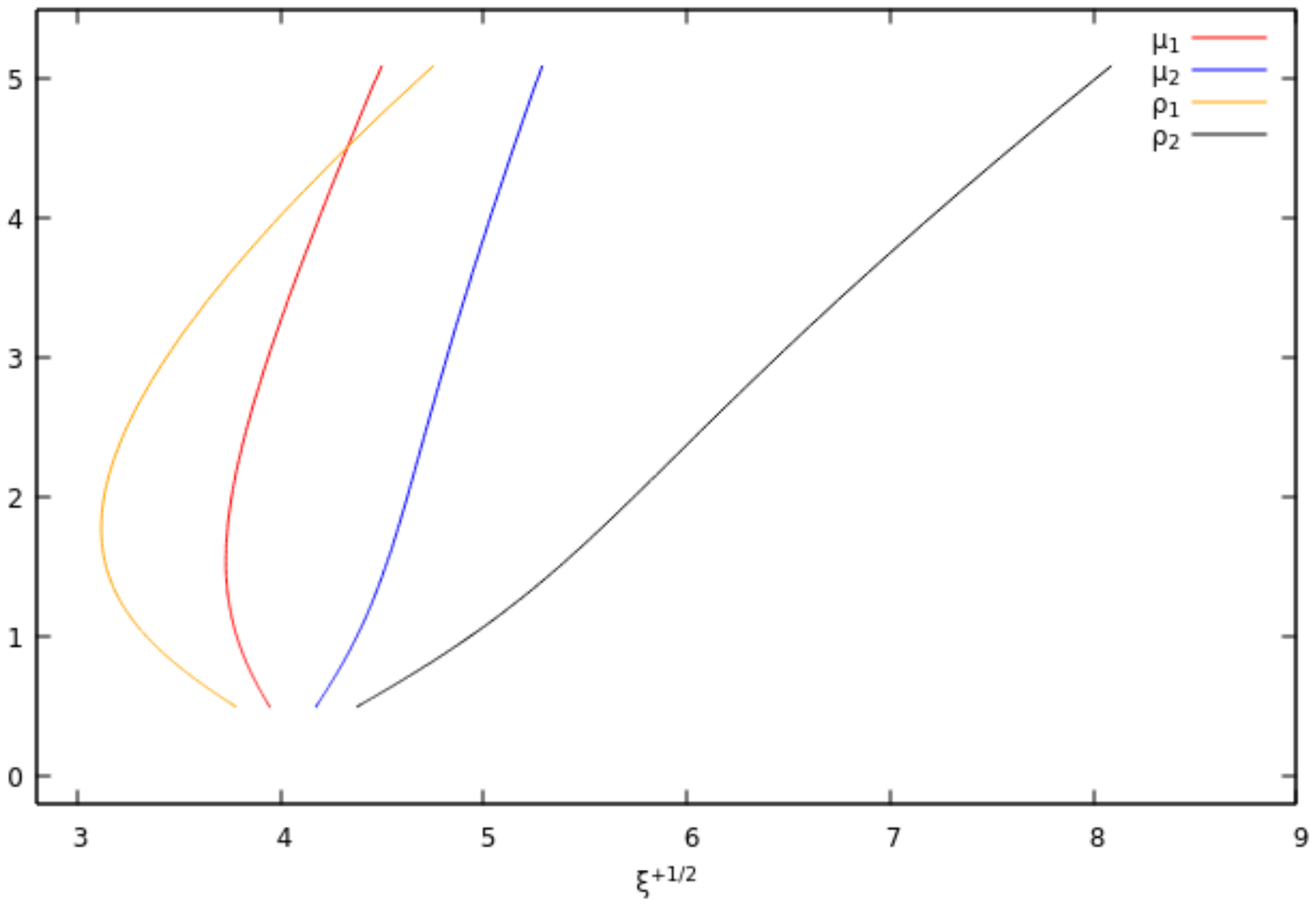}
\vspace{-3cm}
\caption{We show the chemical potentials $\mu_i$ and charge densities $\rho_i$, $i=1,2$, respectively, as functions  of the condensate value $\varphi_+$ (left) and $\xi_+$ (right) for the unbalanced case with $\epsilon=-0.9$ and $\lambda=0.0$.}
\label{fig:unbalanced_epsm09} 
\end{figure}

\begin{figure}[h]
\vspace{-3cm}
\includegraphics[scale=0.4]{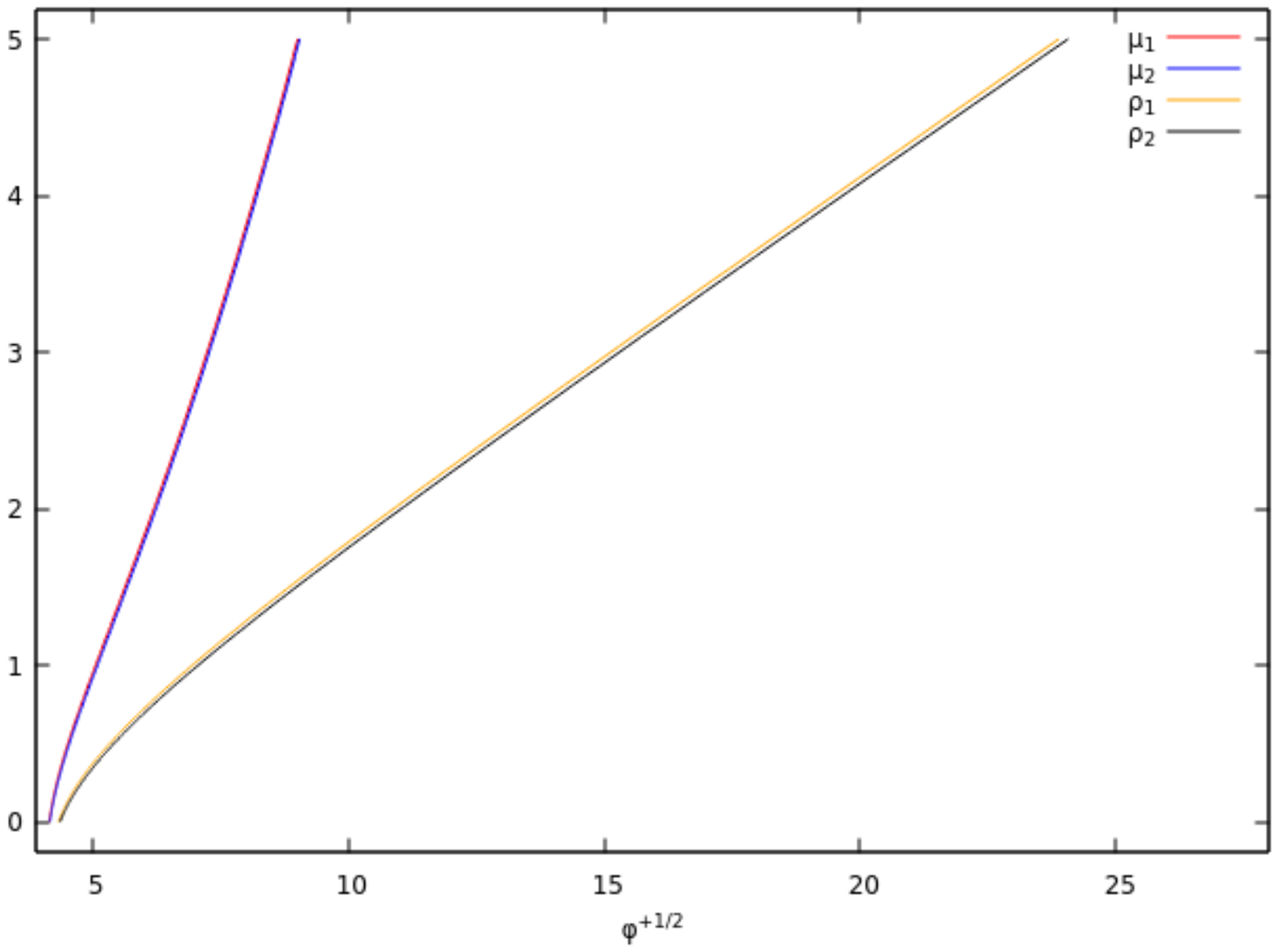}
\includegraphics[scale=0.4]{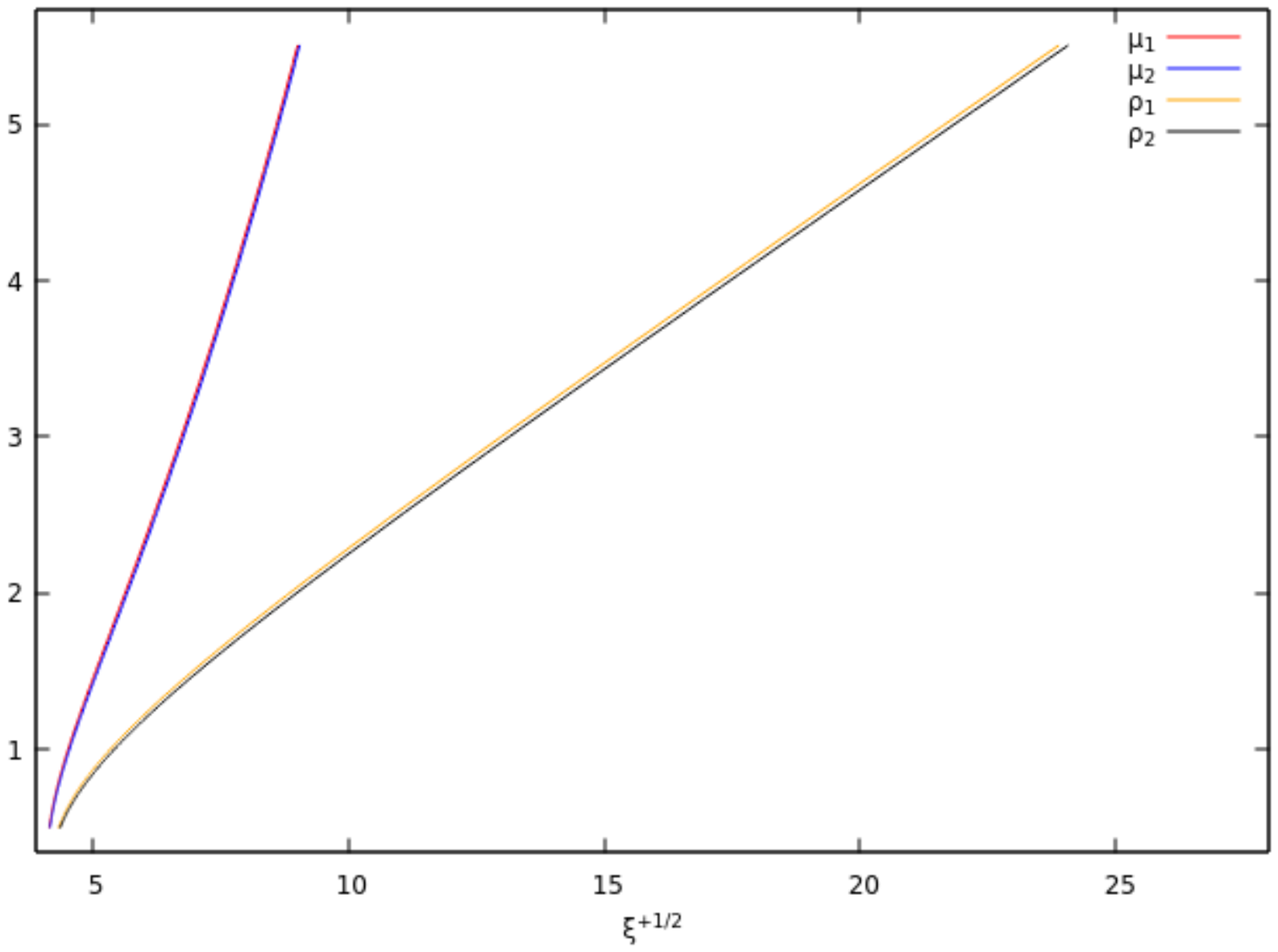}
\vspace{-3cm}
\caption{We show the chemical potentials $\mu_i$ and charge densities $\rho_i$, $i=1,2$, respectively, as functions of the condensate value $\varphi_+$ (left) and $\xi_+$ (right) for the unbalanced case with $\epsilon=0.9$ and $\lambda=0.0$}
\label{fig:unbalanced_eps09} 
\end{figure}

Let us now discuss the unbalanced case which shows some interesting new features.
The most interesting question here is how the chemical potentials and charge densities change when increasing the condensation values $\varphi_+$ and $\xi_+$, respectively. 
We can see that $\epsilon$ greatly influences the charge densities and the chemical potentials of the system. 
When $\epsilon$ takes values close to its negative lower bound,
the inhomogeneous characteristic of the system prevails (see Fig.~\ref{fig:unbalanced_epsm09}), but when $\epsilon$ takes values close to its positive upper bound
(see Fig.~\ref{fig:unbalanced_eps09}) its dominance makes the unbalanced system almost approach a balanced system so that $\rho_1\simeq\rho_2$ and $\mu_1\simeq\mu_2$.

\subsubsection{LOFF phase}

\begin{figure}[h]
\vspace{-1cm}
\includegraphics[scale=0.3, angle=-90]{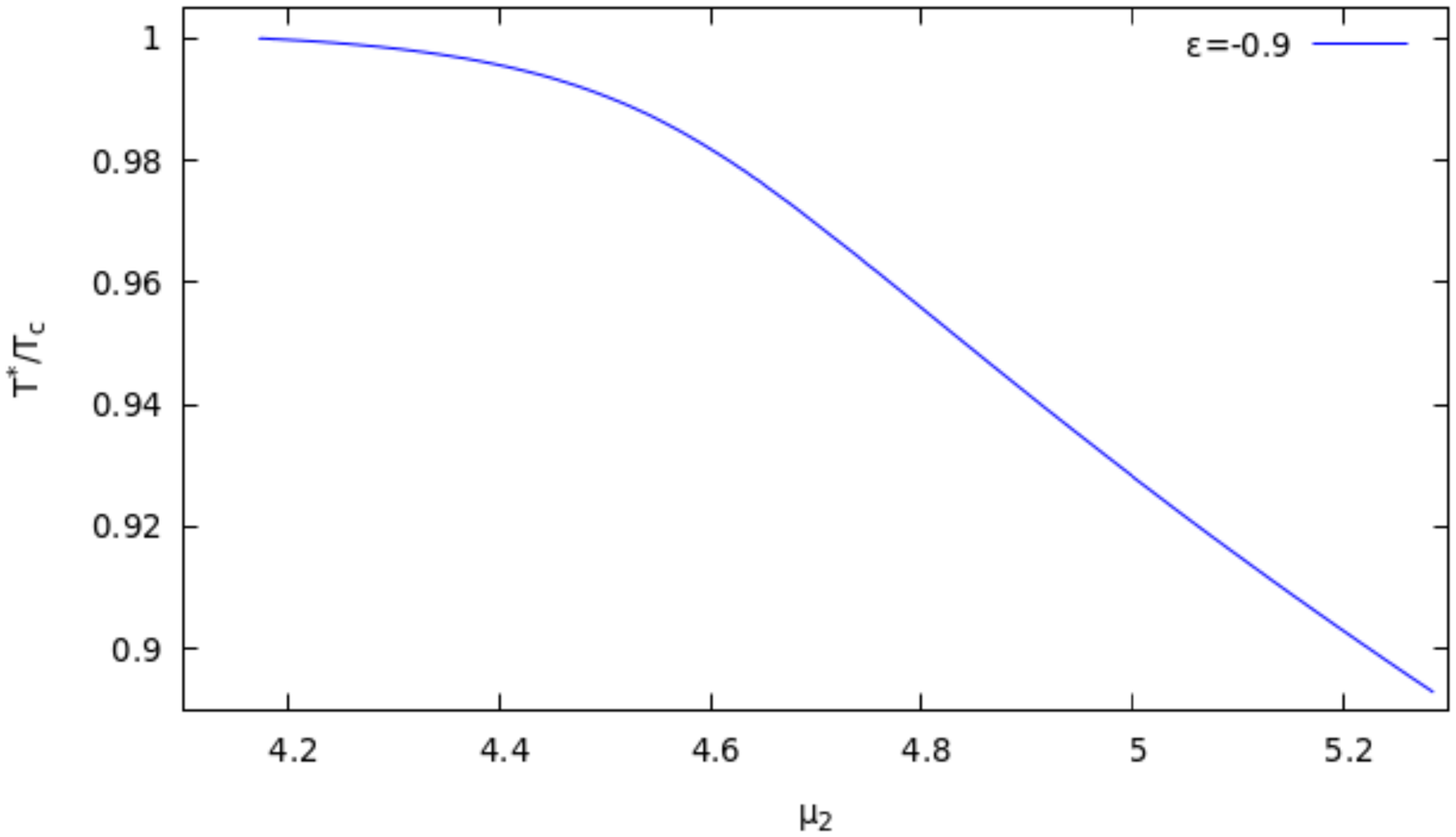}
\includegraphics[scale=0.3, angle=-90]{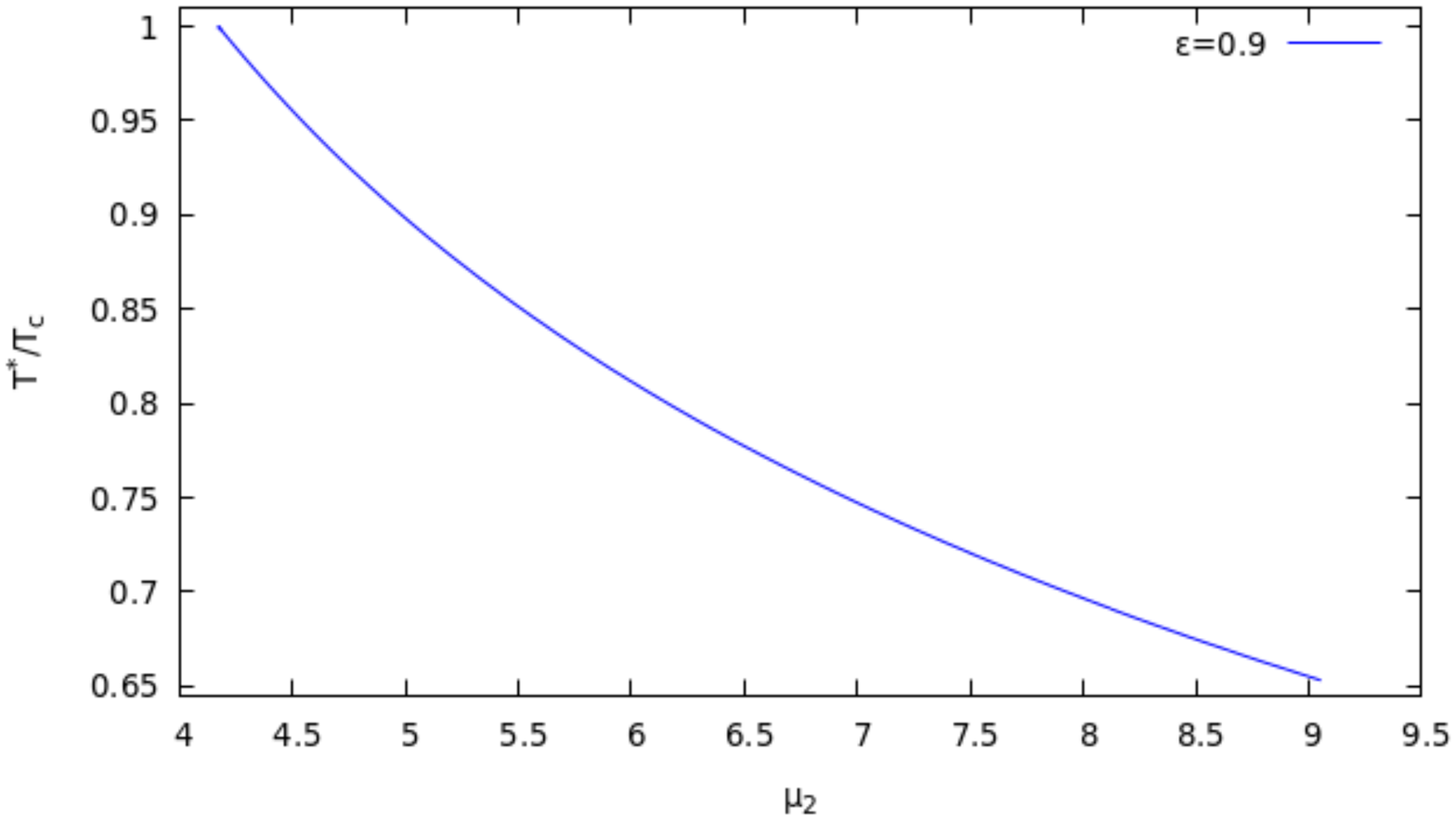}
\vspace{-1cm}
\caption{Unbalanced superconductors with $\epsilon=-0.9$ (left) and $\epsilon=0.9$ (right), respectively, in the $(T/T_c, \mu_2)$ plane. 
The curves demonstrate that there is no sign of the formation of a LOFF phase.}
\label{fig_loof2}
 \end{figure}

As discussed above, in favorable situations such as cuprate-based superconductors, whose superconductivity occurs in planes, the application of an external magnetic field to the superconducting material allows the Zeeman interaction of this field with the electronic spin have a dominant character, so that a situation can be created in which an unbalanced chemical potential $\delta\mu$ appears in the system and can be identified as the application of the field itself ($\delta \mu=H_{z }\mu_{B}$). 
If we allow in our model that the second gauge field $a_\mu$ is interpreted as this field applied to the system, we can associate $\delta \mu=\mu_2$. 
The question then is whether in this situation of non-homogeneous superconductivity the spontaneous breaking of symmetries in the system could favor the emergence of the LOFF phase. 
Our results are given in Fig.~\ref{fig_loof2} for $\lambda=0$ and $\epsilon=-0.9$ (left) and $\epsilon=0.9$ (right).

Note that $\mu_2$ in Fig.~\ref{fig_loof2} represents the unbalanced chemical potential. The curves stop here due to numerical difficulties when increasing
the condensation values further. However, the qualitative form of the curves suggests that the critical temperature never becomes zero 
for finite values of $\mu_2$. If it became zero, this would signal the existence of the LOFF phase. However, as our results suggest, in the set-up
described in this paper, we have not been able to identify it.

\section{Summary}

The interesting thing about materials that belong to the broader class called correlated electron materials such as ferromagnetic unconventional superconductors is that we can promote the tuning between two or more phases by adjusting some parameters external to the system, such as pressure, doping and even temperature. The corresponding phase transitions present in these materials possess typically a quantum critical point and the system exhibits a scale invariance, such that the main characteristics of the phase transitions are preserved over the long range. 
Hence, a better understanding of the dynamics of quantum criticality, as well as the understanding of its behavior with parameters of multiple orders becomes necessary \cite{sachdev2011quantum}. 
In this sense, holography can play a fundamental role.

In this paper, we have introduced a new type of interaction between the two order parameters of the model. This interaction can be interpreted as a direct coupling between the electric and spin motive forces acting on the charge carriers. We find that in comparison to the interaction mediated by a coupling in the potential
of the scalar fields this direct coupling has a strong influence on the quantitative and qualitative behaviour of the system. We find e.g. when the coupling parameter associated to the new interaction is large enough that the value of the order parameter as a  
function of the temperature is practically independent of the coupling constant in the scalar potential.
Here, we have only investigated a
small range of parameters and studied the system only in the probe limit, but other parameter ranges and the study of the system away from the probe limit will be discussed in a future publication.

\section*{Acknowledgement}

We gratefully acknowledge support by the
DFG Research Training Group 1620 {\sl Models of Gravity}. N.P.A.~acknowledges support by CAPES. J.K.~also acknowledges networking support by the COST Actions CA15117 {\sl CANTATA} and CA16104 {\sl GWverse}.

\end{document}